# Steady-state photoconductivity and multi-particle interactions in rubrene single crystals.

P. Irkhin[1], H. Najafov[1] and V. Podzorov[1,2]

Sept. 20, 2014

[1] Department of Physics and Astronomy, [2] Institute for Advanced Materials and Devices for Nanotechnology (*IAMDN*), Rutgers University, Piscataway, NJ 08854, USA

(Corr. author: podzorov@physics.rutgers.edu)

**Abstract:**

We demonstrate that photoconductivity of pristine rubrene crystals exhibits several distinct regimes, in which photocurrent as a function of *cw (continuous wave)* excitation intensity is described by a power law with exponents sequentially taking values 1, 1/3 and ¼. We show that this photocurrent is generated almost exclusively at the surface of pristine rubrene crystals, while the bulk photocurrent is dramatically smaller and follows a different set of exponents, 1 and ½. A model based on exciton fission, fusion and triplet-charge quenching is developed to describe these non-trivial effects in photoconductivity of highly ordered organic semiconductors.

Future advances in organic electronics will require better fundamental understanding of the intrinsic charge transport and optical properties of organic semiconductors that can be efficiently gained by investigating organic single-crystal devices, where disorder is minimized. Among the most important processes are the polaronic charge transport [1,2] and dynamics of singlet and triplet photoexcitations, the excitons [3,4]. Rubrene (5,6,11,12-tetraphenylnaphthacene) single crystals attracted much attention due to many compelling properties, including nearly trap-free charge transport [5], one of the highest charge carrier mobilities of up to 20 $cm^2V^{-1}s^{-1}$ [5,6,7,8], high photoconductivity [9,10] and large triplet exciton diffusion length [9,11], making this system ideal for fundamental studies. Here, we investigate steady-state photoconductivity in rubrene and report several non-trivial regimes arising from multi-particle interactions that involve singlet and triplet excitons, as well as charge carriers. Many of these processes depend on the probability of one particle to find another and thus become important in systems, where excitons and charge carriers are sufficiently mobile and have long life-times.

We use high-purity vapor grown rubrene single crystals (for details see, e.g., [12]) with silver or carbon contacts deposited at the (*a*,*b*) facet, separated by 25 μm to ~ 3 mm (Fig. 1). The (*a*,*b*) facet is uniformly illuminated at normal incidence with a monochromatic light in the highly-absorbing energy range of rubrene [13], and photoconductivity, $\sigma_{PC}$, and photoluminescence (PL) are measured simultaneously. All the measurements were performed in small applied electric fields, in the range 10 - 100 V·$cm^{-1}$, where all the reported effects were found to be independent of this field. A special care has been taken to make sure the crystals do not degrade under photoexcitation in the entire excitation power range. Most measurements were performed in a clean high vacuum ($10^{-5}$ Torr) with a high-vacuum gauge turned off,unless it was used intentionally, as described below. Optical excitation power (*P*, in nW) refers to the total incident power measured at the surface of the sample. Irradiance, or excitation intensity ($\Phi$, in nW/cm$^2$) refers to the excitation power per unit of illuminated area of the sample. We also plot our data as a function of the density of absorbed photons ($G$, in cm$^{-3}$s$^{-1}$) defined as the number of photons absorbed per cm$^3$ per second at the illuminated surface of the crystal: $G = \Phi/(h\nu \cdot \alpha^{-1})$, where $h\nu$ is the excitation photon energy, and $\alpha$ is the absorption coefficient of crystalline rubrene (not to be confused with the power exponent α) [13]. For further technical details, see Supplementary Information, sec. 1.

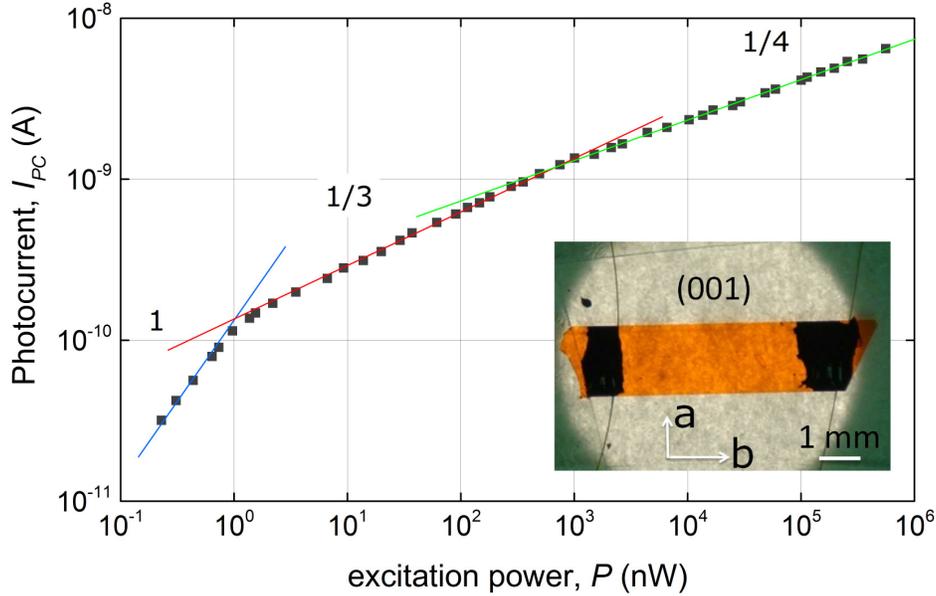

**Fig. 1. Dependence of the steady-state photocurrent on the excitation power in pristine rubrene.** Conditions: λ = 500 nm, normal incidence at (***a***,***b***) facet, *V* = 1 V. Note that excitation intensity is varied over 6 orders of magnitude. Inset: a photograph showing the typical sample geometry. Photocurrent follows a power law, $\sigma_{PC}(G) \propto G^\alpha$, with the three distinct regimes described by the power exponent α = 1, 1/3 and ¼ (note the double-log scale).

Figure 1 shows the typical result of photoconductivity measurements in pristine rubrene performed in a wide range of excitation intensities covering more than six decades in light power. Pure photocurrent $I_{PC}$ or photoconductivity $\sigma_{PC}$ are defined as the total current or conductivity under illumination minus the dark current or conductivity (all measured at a small voltage, $V$, applied between the contacts). The most striking observation is that photoconductivity closely follows a power law, $\sigma_{PC} \propto G^{\alpha}$, with the exponent $\alpha$ sequentially taking values 1, 1/3 and ¼, as the excitation power increases. We reproducibly observe these three distinct regimes in pristine crystals.

Before continuing with the investigation of this non-trivial photoconductivity, we will first demonstrate that most of the photocurrent in pristine rubrene crystals flows at the very surface of the crystal, perhaps within a ~ nm below the (*a*,*b*) surface, while bulk photocurrent is contributing less than ~ 1% to the total photoconductivity. This very surprising effect can be convincingly demonstrated by using the so-called "*gauge effect*". Previously, it was discovered that high-vacuum gauges generate electrically neutral free radicals in high-vacuum chambers that land at exposed surfaces of molecular crystals and create traps, leading to a reduced charge carrier mobility and increased threshold voltage in vacuum-gap OFETs [14]. Here, we show that high-vacuum gauges also drastically reduce photoconductivity (Fig. 2). In this measurement, dark and photoconductivity of a pristine rubrene crystal ($\sigma_0$ ~ 1 nS and $\sigma_{PC}$ ~ 3.7 nS, respectively) were monitored in high vacuum (with the gauge off). At $t = 250$ s, a high-vacuum gauge was turned on, which resulted in a very fast and drastic decrease of both $\sigma_0$ and $\sigma_{PC}$. The very small photoconductivity pulses at $t > 250$ s are produced by the same illumination as the pulses at $t < 250$ s. The inset in Fig. 2 shows the "gauge effect" on the pure photoconductivity, $\sigma_{PC}$, calculated by subtracting the dark conductivity background, $\sigma_0$, from the total conductivity under illumination, $\sigma$. Remarkably, while the dark conductivity is reduced by the gauge by a factor of 20, the photoconductivity is diminished by a staggering factor of 60. For more details on the dark surface conductivity in rubrene see Supplementary Information, sec. 2.

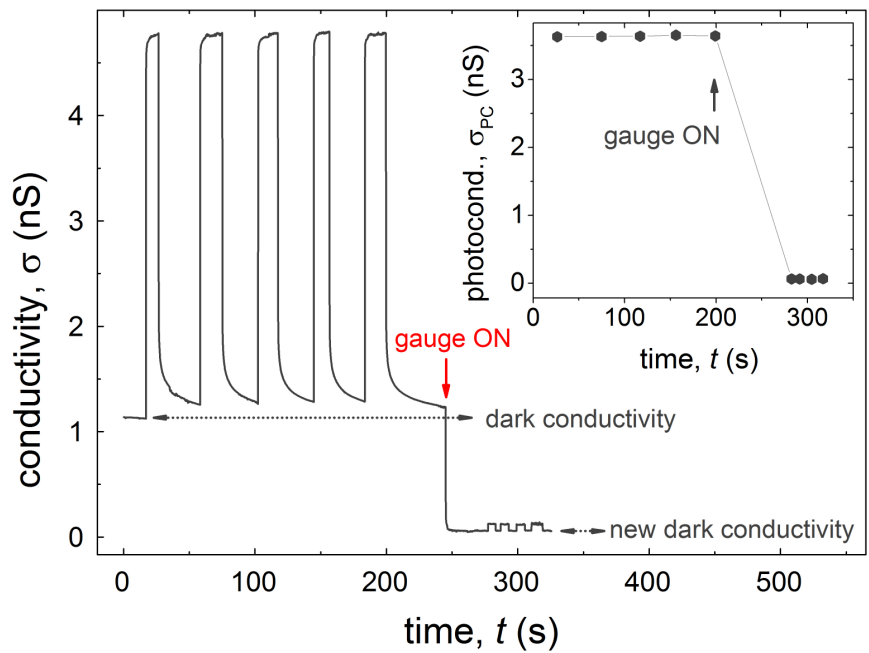

**Fig. 2. Demonstration of the "gauge effect" on dark and photoconductivity in pristine rubrene.** Total conductivity, $\sigma(t)$, is monitored in high vacuum ($10^{-5}$ Torr) with the high-vacuum gauge off ($t < 250$ s), and after the gauge was turned on ($t > 250$ s). The pulses in $\sigma$ correspond to photoexcitation with square light pulses ($\lambda = 532$ nm, normal incidence to ($a,b$) facet, $P = 8$ μW, pulse duration 10-20 s, $V = 20$ V). The same photoexcitation pulses were used before and after the gauge was turned on. The inset shows the pure photoconductivity, $\sigma_{PC}$ (with the dark background subtracted), obtained from each pulse. It is clear that the operating high-vacuum gauge diminishes photoconductivity by almost 100%.

As discussed in Ref. 14, "gauge effect", which is a short-range effect, creates charge traps at organic surfaces, thus having a strong influence on the *surface* conductivity of the crystals. Photoexcitation used in the above experiment in Fig. 2 has a large penetration length ($\alpha^{-1} = 10$ μm) and is thus generating a distribution of excitons in the bulk of the crystal. In a conventional "bulk" model of photoconductivity, photocarriers are created locally, where photons are absorbed, thus leading to samples photoconducting throughout the bulk. Therefore, if conventional model of photoconductivity were applicable, gauge effect would be expected to result only in a minor decrease of $\sigma_{PC}$, because of the short-range nature of charge carrier interactions with neutral free radicals at the crystal surface. A dramatic (nearly 100%) reduction of $\sigma_{PC}$ with the gauge effect in our experiment thus suggests that photoconductivity in pristine rubrene is predominantly a surface effect. The remaining small photocurrent, flowing after the gauge-effect treatment of the crystal, can now be attributed to a leftover bulk photoconductivity.

This remarkable result alone lends a strong support to the model, in which photon absorption leads to a generation of long-lived excitonic species (for rubrene, triplets generated via singlet fission) capable of reaching the surface of the crystal from the bulk due to long-range diffusion, where they dissociate (with a certain quantum efficiency) and generate a surface photocurrent. Irrespectively of the detailed microscopic mechanism of triplet exciton dissociation at the surface, the gauge effect gives us a powerful tool to selectively trap otherwise mobile carriers only at the surface of the crystal and switch the sample from a surface to a bulk photoconductor, without morphologically or chemically damaging the surface. In addition, such a transition is reversible, because the gauge effect damage can be fully recovered [14]. Thus, we can measure photoconductivity and its photoexcitation intensity dependence, $\sigma_{PC}(G)$, in either the surface or the bulk regimes independently. Note that gauge effect should not noticeably affect the exciton dissociation rate at the surface because of the typically low densities ($10^9$ - $10^{12}$ cm$^{-2}$) of generated traps (Supplementary Information, sec. 2) [14]. Below we show that these dependencies are remarkably different.

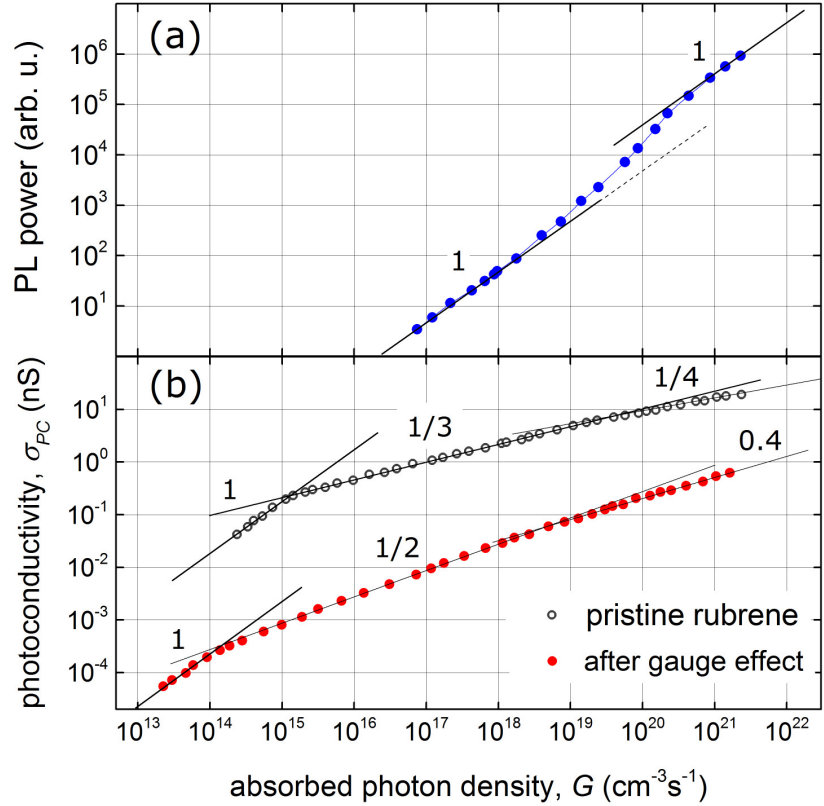

**Fig. 3. Photoluminescence (a) and photoconductivity (b) simultaneously measured in rubrene crystals as a function of excitation density.** Photoconductivity is shown for pristine crystal (black open circles), and for the same crystal after it has been exposed to a high-vacuum gauge (red solid circles). Thin solid lines are power-law fits, $\sigma_{PC} \propto G^{\alpha}$, with the exponent $\alpha$ indicated next to each line (note the double-log scale). PL was similar before and after applying the gauge.

After the relative contributions of the surface and bulk photoconductivities have been established, we focus our attention on the two limiting cases, the surface-dominated and bulk-dominated photoconductivity, measured in the same crystal in its pristine and gauge-treated states, respectively (Fig. 3b). In addition, photoluminescence power emitted from the (*a*,*b*)-facet of the crystal has been measured simultaneously with photoconductivity (Fig. 3a). The gauge effect reduces the absolute value of $\sigma_{PC}$ by 1-2 orders of magnitude, and, more interestingly, leads to a qualitative change in the excitation intensity dependence of photoconductivity from a power-law exponent $\alpha = 1/3$ (surface-dominated) to $\alpha = ½$ (bulk-dominated). It is interesting that the exact method of introducing surface defects is not important: we observed the same effect of reduction of $\sigma_{PC}$ and a conversion from $\alpha = 1/3$ to $\alpha = ½$ after intentional photooxidation of (*a,b*) facets or rubrene, which is known to introduce traps near the surface (Supplementary Information, sec. 3) [15].

Photoconductivity of pristine crystal (black open symbols in Fig. 3b) exhibits the three well-defined regimes characterized by the power-law exponents $\alpha = 1$, $1/3$ and ¼. The bulk photoconductivity of the same crystal, obtained by exposing it to a high-vacuum gauge, has a different behavior with the transitions from $\alpha = 1$ to ½ and 0.4 (red solid symbols in Fig. 3b). On the lowest intensity end (below $10^{14}$ - $10^{15}$ cm$^{-3}$ s$^{-1}$), the simple linear dependence is always observed in our crystals. At high excitation powers, a significant increase of PL yield is observed (a "bump" in PL in Fig. 3a), with the inflection point at around $10^{20}$ cm$^{-3}$ s$^{-1}$, signifying a transition from one linear regime to another one with ~ 10 times greater slope. This crossover in PL approximately coincides with the transition from $\alpha = 1/3$ to ¼ in the surface photoconductivity of pristine crystals. Such a drastic increase of photoluminescence quantum yield thus correlates well with a reduction in the photocarrier generation efficiency, implying that the same new channel responsible for formation of additional emissive singlets is also responsible for an extra photocarrier loss. This correlation and its implications are discussed in more detail below.

The model formulated below rationalizes our experimental observations. The initial photoexcitation in this class of materials results in a short-lived singlet molecular exciton [3,4]. In this work, we simultaneously detect PL emitted upon radiative recombination of singlets and the excess electrical conductivity (the photoconductivity) that arises when excitons dissociate, thus taking advantage of complementarity of PC and PL. Several recent experimental results pointed towards the possibility that singlet excitons in rubrene efficiently generate triplets via singlet fission, presumably because singlets have energy that is nearly twice the energy of the triplets [16]. Singlet fission competes with a direct radiative relaxation of singlets. In addition, two triplet excitons can recombine via triplet fusion to produce an emissive singlet that has the same characteristics as the initially photoexcited singlet. Additional consideration must be added in the case of rubrene (or, other highly ordered small-molecule semiconductors, such as tetracene or pentacene), because the triplet excitons in these material are mobile [9,11,17], meaning that the probability that they can interact with each other and undergo fusion is high. The radiative recombination of triplets is quantum-mechanically forbidden, resulting in a typical triplet lifetime (100 µs in rubrene crystals) to be much longer than the lifetime of singlets (a few ns) [16]. This in turn allows for the accumulation of large concentrations of triplets under steady-state excitation conditions, and triplets can start to interact, for instance, with available free carriers: $T_1 + n \rightarrow S_0 + n$, which has a spin-allowed channel, or with each other. The possibilities of triplet-charge quenching and triplet fusion have been recently discussed theoretically and experimentally in the context of bulk-heterojunction solar cells and fluorescence measurements [18,19,20]. In addition, because of their long lifetime and diffusion range, triplet excitons are the most feasible candidates for reaching dissociation sites (such as surfaces and interfaces) and contributing to photoconductivity.

We use a set of rate equations to describe the dynamics of the charge carrier density ($n$), and the density of singlet ($S$) and triplet ($T$) excitons, in the presence of singlet-to-triplet and triplet-to-singlet conversions occurring via fission and fusion, as well as *triplet-carrier quenching*:

$$\frac{dS}{dt} = G - \frac{S}{\tau_S} + \frac{1}{2} f_T \gamma_T T^2 \quad (1)$$

$$\frac{dT}{dt} = -\frac{T}{\tau_T} + 2 f_S \frac{S}{\tau_S} - \gamma_T T^2 - \beta T - \gamma_a T n \quad (2)$$

$$\frac{dn}{dt} = \beta T - \gamma_n (n_0 + n) n \quad (3)$$

Here, $G$ is a singlet exciton generation term via light absorption, $n_0$ is the dark carrier concentration (corresponding to the built-in surface conduction channel existing in pristine rubrene [14]), $f_S$ is the probability that singlet decay results in fission into two triples, $f_T$ is the probability that triplet-triplet collision results in the creation of a singlet via fusion, $\tau_S$ and $\tau_T$ are the singlet and triplet exciton lifetimes, respectively, $\gamma_T$ is a bimolecular recombination coefficient for triplets, and $\gamma_n$ is a bimolecular recombination coefficient for charge carriers. Factors 2 and ½ take into account the fact that fission of one singlet results in two triplets, and vice versa. An additional term that appears on the right hand side of eq. 2 (proportional to $T \cdot n$) allows for annihilation of triplet excitons via collision with charge carriers, with the corresponding probability $\gamma_a$.

We also take into account the conclusion reached in this work that most of the photocurrent in pristine rubrene crystals is produced via dissociation of triplets at the surface of the crystal. A possible mechanism may involve interactions of triplets with surface defects. However, given the clean, oxide-free surfaces of pristine rubrene crystals recently confirmed in the surface analytical studies [21], an intrinsic dissociation effect due to the lattice discontinuity is more likely. Such dissociation, occurring with probability $\beta$, results in a population of photocarriers, $n$, responsible for photoconductivity $\sigma_{PC} = en\mu$, where e is the elementary charge, and $\mu$ is the charge carrier mobility. These charge carriers have a lifetime much longer than that of singlet or triplet excitons. In a dynamic equilibrium corresponding to *cw* excitation conditions in our experiment, the densities of all the involved species are constant: $dS/dt = dT/dt = dn/dt = 0$.

In the limit of weak photoexcitations, the terms in eq. 1-3 corresponding to multi-particle interactions can be neglected ($T^2$, $T \cdot n \approx 0$) because of the low densities of these species, the density of photogenerated carriers is low ($n \ll n_0$), and the concentrations of singlets, triplets and photocarriers are proportional to photoexcitation intensity: $S, T, n \propto G$, which corresponds to the linear ($\alpha = 1$) regime in surface and bulk photoconductivity. At somewhat higher excitation intensities, when triplet-triplet and triplet-charge terms are still very small and the generation of singlet and triplet excitons is still linear ($S, T \propto G$, from eqs. 1-2), the carrier concentration can be sufficiently high ($n \gg n_0$) for a bimolecular electron-hole recombination to dominate, resulting in a $n \propto G^{1/2}$ dependence for the bulk photoconductivity (eq. 3), a typical result of carrier recombination in high-mobility semiconductors. The crucial difference with rubrene is that we only observe this regime in the bulk, after the surface photoconductivity is eliminated by subjecting the crystal to the gauge effect creating surface traps (Fig. 3b).

The reason bimolecular recombination ($\alpha = 1/2$ regime) is not observed in pristine crystals is that in the case of surface-dominated photoconductivity, the concentration of mobile photocarriers and triplet excitons accumulated at the surface can get much higher and increase much faster with $G$ than that in the bulk, and interactions between triplet excitons and charge carriers become dominant even at moderate photoexcitation intensities. In this regime, triplet fusion is still negligible ($T^2 \sim 0$, $S \propto G$ in eq. 1), and there are only two dominant terms in eq. 2: the triplet generation via singlet fission ($2f_S S/\tau_S$) and the triplet-charge quenching ($\gamma_a T \cdot n$), leading to the following relationship between the concentrations of triplets and photocarriers near the surface: $T \cdot n \propto G$. Thus, in this regime, we get for the carrier density from eq. 3: $n \propto G^{1/3}$. It must be noted that although a regime with $\alpha = 1/3$ can stem from 3$^{rd}$ order (three particle) Auger recombination processes, such processes in rubrene are unlikely, as higher order effects usually occur in indirect gap inorganic semiconductors, where radiative recombination is prohibited, and they require much greater excitation densities (Supplementary Information, sec. 4).

At the highest excitation densities ($\alpha = \frac{1}{4}$ for the surface and 0.4 for bulk-dominated photoconductivity), triplet fusion represented by $T^2$ term in eqs. 1 and 2 kicks in, strongly competing with all other triplet decay channels, thus leading to $T \propto S^{1/2}$ (from eq. 2) and $n \propto G^{1/4}$ (from eq. 3) dependences. Note that the population of singlets $S$ remains linear with $G$ even in this case, although with a much greater slope. Indeed, the solution of eq. 1 without and with the triplet fusion term gives $S = \tau_S G$ and $S = \tau_S G/(1 - f_S f_T)$, respectively. This indicates that the power dependence of PL remains linear in both regimes, but the slope increases by a factor of $1/(1 - f_S f_T)$ when triplet fusion becomes dominant. In rubrene, we systematically observe a significant (a factor of ~ 10) increase in photoluminescence yield at around $10^{20}$ cm$^{-3}$ s$^{-1}$ (Fig. 3a). This suggests that the product $f_S f_T \approx 0.9$, and thus the probabilities of singlet fission and triplet fusion are individually very high ($f_S \geq 0.9$ and $f_T \geq 0.9$), approaching ~ 100%. This shows that in rubrene both fission and fusion processes are extremely efficient. Note that $f_S$ and $f_T$ rates are the probabilities for individual excitons to undergo fission or fusion, respectively. The remaining ~ 10% of the probability in this excitation regime corresponds to other (non-dominant) channels of exciton decay, such as a radiative recombination for singlets, triplet-charge quenching and triplet dissociation. All these processes are self-consistently accounted for in the system of equations (1-3). Recent time resolved and steady-state spectroscopic measurements [16,[22],[23]], as well as pump-probe studies [[24]], provide more information on exciton fission and fusion dynamics in rubrene.

Finally, it must be noted that in light of the new knowledge of non-linear photoconductivity in rubrene observed here, the estimate for the exciton diffusion length, $L_{EX}$, previously reported by Najafov *et al*. [9] must be adjusted. The refined procedure should incorporate the 1/3 power dependence in the modeling of the photocurrent modulations, $\sigma_{PC}(\lambda,\theta)$, with excitation polarization angle $\theta$ and wavelength $\lambda$, and it would thus result in somewhat different exciton diffusion lengths. The detailed calculations show that $L_{EX}$ remains well above a micrometer ($L_{EX} > 1$ µm) for typical photocurrent oscillations observed in high-quality pristine rubrene crystals (see Supplementary Information, sec. 5, for the refined model).

To summarize, we have shown that contrary to the common expectation of a linear photo-response, highly-ordered organic semiconductors can exhibit a non-linear excitation density dependence of photoconductivity in steady-state measurements. We have devised a simple, yet powerful, experimental method for separating bulk from surface photoconductivity, based on gauge effect. It allowed us for the first time to observe that surface photoconductivity in pristine triplet organic semiconductors, such as rubrene, follows a power law with exponents 1, 1/3 and ¼, while bulk photoconductivity follows the more conventional exponents 1 and ½. We have developed a model based on exciton fission and fusion, as well as triplet-charge quenching, that describes this non-trivial behavior. The phenomenological and fundamental understanding of strong nonlinearities in photoconductivity provided by this work is important for gaining deeper insights into the physics of excitons and charge carriers in organic semiconductors and utilization of these materials in emerging photonic and electronic applications.

**Acknowledgement.** Acknowledgement is made to the donors of the American Chemical Society Petroleum Research Fund for support of this research (grant # 50629-DNI10). We are indebted to Dr. Yuanzhen Chen for his help with measurements, as well as for useful discussions. We are grateful to Szu-Ying Wang for her help with the rubrene crystal growth. Finally, we thank Profs. V. M. Agranovich, Yu. Gartstein and H. Bässler for helpful theoretical insights.

**References:**

[1] V. Coropceanu, Y. Li, Y. Yi, L. Zhu and J.-L. Bredas, MRS Bull. **38**, 57 (2013).

[2] D. L. Cheung and A. Troisi, Phys. Chem. Chem. Phys. **10**, 5941 (2008).

[3] C. J. Bardeen, MRS Bull. **38**, 65 (2013).

[4] J. R. Tritsch, W.-L. Chan, X. Wu, N. R. Monahan and X.-Y. Zhu, Nature Comm. **4**, 2679 (2013).

[5] V. Podzorov, E. Menard, A. Borissov, V. Kiryukhin, J. A. Rogers, and M. E. Gershenson, Phys. Rev. Lett. **93**, 086602 (2004).

[6] Y. Xia, J. H. Cho, J. Lee, P. P. Ruden, and C. D. Frisbie, Adv. Mater. **21**, 1 (2009).

[7] W. Xie, K. A. McGarry, F. Liu, Y. Wu, P. P. Ruden, C. J. Douglas, and C. D. Frisbie, J. Phys. Chem. C **117**, 11522 (2013).

[8] Y. Okada, K. Sakai, T. Uemura, Y. Nakazawa, and J. Takeya, Phys. Rev. B **84**, 245308 (2011).

[9] H. Najafov, B. Lee, Q. Zhou, L. C. Feldman, and V. Podzorov, Nature Mater. **9**, 938 (2010).

[10] O. Ostroverkhova, D. G. Kooke, F. A. Hegmann, J. E. Anthony, V. Podzorov, M. E. Gershenson, O. D. Jurchescu, and T. T. M. Palstra, Appl. Phys. Lett. **88**, 162101 (2006).

[11] P. Irkhin and I. Biaggio, Phys. Rev. Lett. **107**, 017402 (2011).

[12] V. Podzorov, MRS Bull. **38**, 15 (2013).

[13] P. Irkhin, A. Ryasnyanskiy, M. Koehler, and I. Biaggio, Phys. Rev. B **86**, 085143 (2012).

[14] V. Podzorov, E. Menard, S. Pereversev, B. Yakshinsky, T. Madey, J. A. Rogers, and M. E. Gershenson, Appl. Phys. Lett. **87**, 093505 (2005).

[15] H. Najafov, D. Mastrogiovanni, E. Garfunkel, L. C. Feldman, and V. Podzorov, Adv. Mater. **23**, 981 (2011).

[16] A. Ryasnyanskiy and I. Biaggio, Phys. Rev. B **84,** 193203 (2011).

[17] G. M. Akselrod, P. B. Deotare, N. J. Thompson, J. Lee, W. A. Tisdale, M. A. Baldo, V. M. Menon and V. Bulovic, Nature Comm. **5**, 3646 (2014).

[18] A. D. Poletayev, J. Clark, M. W. B. Wilson, A. Rao, Y. Makino, S. Hotta, and R. H. Friend, Adv. Mater., **26**, 919 (2014),

[19] N. J. Thompson, E. Hontz, D. N. Congreve, M. E. Bahlke, S. Reineke, T. Van Voorhis, and M. A. Baldo, Adv. Mater. **26**, 1366 (2014)

[20] M. Wittmer and I. Zschockkle-Gränacher, J. Chem. Phys. **63**, 4187 (1975).

[21] D. D. T. Mastrogiovanni, J. Mayer, A. S. Wan, A. Vishnyakov, A. V. Neimark, V. Podzorov, L. C. Feldman, and E. Garfunkel, Sci. Reports **4**, 4753 (2014).

[22] S. Tao, N. Ohtani, R. Uchida, T. Miyamoto, Y. Matsui, H. Yada, H. Uemura, H. Matsuzaki, T. Uemura, J. Takeya, and H. Okamoto, Phys. Rev. Lett. **109**, 097403 (2012).

[23] I. Biaggio and P. Irkhin, Appl. Phys. Lett. **103**, 263301 (2013).

[24] L. Ma, K. Zhang, C. Kloc, H. Sun, M. E. Michel-Beyerle, and G. G. Gurzadyan, Phys. Chem. Chem. Phys. **14**, 8307 (2012).

**Steady-state photoconductivity and multi-particle interactions in rubrene single crystals.**

(Sept. 20, 2014)

P. Irkhin[1], H. Najafov[1] and V. Podzorov[1,2]

[1] Department of Physics and Astronomy, [2] Institute for Advanced Materials and Devices for Nanotechnology (*IAMDN*), Rutgers University, Piscataway, NJ 08854, USA

[*]Corresponding author e-mail address: podzorov@physics.rutgers.edu

## 1. Experimental methods: sample preparation, measurements and reproducibility.

In this work, we use high-purity, pristine rubrene single crystals grown using an optimized physical vapor transport method (for details see, e.g., [1]). Electrical contacts were prepared at the (*a*,*b*) facet of the crystal either by thermally evaporating silver through a shadow mask in high vacuum or by depositing colloidal graphite from an aqueous solution. The contacts are formed in a coplanar configuration (on top of the (*a*,*b*) facet of the crystal) with inter-contact separation ranging from 25 µm to ~ 3 mm. In the typical geometry of our measurements, the (*a*,*b*) facet of the crystal is uniformly illuminated at a normal incidence with a monochromatic light, and photoconductivity, $\sigma_{PC}$, is measured at the (*a*,*b*) facet at room temperature by using Keithley K2400 source-meters and K6512 electrometers. All measurements are carried out in small applied electric fields, below 100 V·cm$^{-1}$. Photocurrent ($I_{PC}$) and photoluminescence (*PL*) are excited with a *cw* (continuous wave) laser emission with a photon energy of 2.3 eV (532 nm) or 2.6 eV (473 nm) in the highly-absorbing energy range of rubrene [2], or by monochromatic light produced by a halogen lamp emission sent through a narrow band filter. *PL* generated in the crystal is collected and imaged onto an optical fiber coupled to the Ocean Optics USB2000 spectrometer. A special care has been taken to make sure the crystals do not degrade under photoexcitation during the entire set of measurements and in the entire range of excitation intensities. We used longpass filters with appropriate edge wavelengths to remove the excitation light from the detected *PL* signal. Excitation density was controlled by a set of neutral density filters, while the distribution of light at the sample's surface between the contacts remained constant and very uniform. Most measurements reported here were performed in a clean high vacuum (10$^{-5}$ Torr) with a high-vacuum gauge turned off, unless it was used intentionally, as described in the main text.

It must be noted that at very high excitation intensities, detrimental effects due to local heating, material damage and photooxidation, leading to irreversible changes in the properties of the crystals, may occur. Thus, we took extra precautions to avoid sample degradation under illumination by consistently checking the reproducibility of all the reported data. All the dependencies presented in this work were confirmed to be reproducible, hysteresis free and independent of the measurement history.

## 2. On the nature of dark and photoconductivity.

The appreciable dark conductivity typically observed at the free (*a*,*b*) surface of pristine rubrene crystals (the so-called *built-in conduction channel*) has been originally attributed to the presence of a monolayer of rubrene endo-peroxide at the surface [3]. This assignment has been prompted by the fact that rubrene molecules in solutions are known to readily photooxidize, leading to rapid bleaching of the liquid samples [4]. However, recent detailed studies of crystalline rubrene by the methods of surface analysis, including X-ray photoelectron and secondary ion mass spectroscopies, have conclusively shown that (*a*,*b*) facets of pristine rubrene crystals are almost free from oxygen [5]. This appears to be due to the tight packing of rubrene molecules in the crystal lattice that makes the surface very resilient against oxidation. Indeed, rubrene single crystals stored in air for as long as a few years (mostly in the dark, but occasionally handled under ambient or microscope illumination) have a surface concentration of atomic oxygen as low as 1-3 % of the full monolayer of rubrene endo-peroxide [5]. Thus, the built-in surface conduction channel in pristine rubrene crystals does not appear to be formed as a result of a surface oxidation, but rather occurs via other mechanisms that could be related, for example, to a band bending at the free crystal surface or small relaxation of the surface structure [6]. Similar mechanisms might also be responsible for the surface dissociation of triplet excitons that reach the surface from the bulk via diffusion and generate a substantial surface photocurrent [7].

Irrespectively of the exact microscopic mechanism of dark and photo-conductivity in rubrene, the gauge effect described in the main text gives us a powerful tool to selectively "turn off" the surface and switch the sample from a surface to a bulk (photo)conductor, without morphologically or chemically damaging the surface. In addition, such a transition is reversible, because the gauge effect can be fully recovered (for details, see [8]). Thus, we can intentionally switch between the two distinct regimes of photoconductivity of the same sample and measure the photoexcitation intensity dependence, $\sigma_{PC}(G)$, of either the surface or the bulk photocurrents independently.

We believe that gauge effect does not significantly influence the triplet dissociation rate at the surface, because the typical density of traps generated by high-vacuum gauges at the surface ($10^9$ - $10^{11}$ cm$^{-2}$) is much smaller than the density of rubrene molecules at the (*a*,*b*) facet of the crystal ($10^{14}$ cm$^{-2}$) [8].

## 3. Photoconductivity and intentional, gradual photooxidation of rubrene crystals.

We have shown in the main text that the excitation density dependence of photoconductivity, $\sigma_{PC}(G)$, in the bulk and surface regimes in rubrene crystals are remarkably different. For the most commonly used photoexcitation intensities, the surface photoconductivity follows a power law, $\sigma_{PC}(G) \propto G^{\alpha}$, with the exponent $\alpha = 1/3$, while the bulk photoconductivity is described by another exponent $\alpha = ½$ (note: this exponent $\alpha$ is not to be confused with the absorption coefficient *α*). We would like to additionally demonstrate that, in principle, all the intermediate states of photoconductivity, between the two limiting regimes of a pure surface (in pristine crystals) and pure bulk (in gauge-effect treated crystals) photoconductivities are possible in trap-dominated crystals. To show this, we intentionally introduced traps in pristine crystals by photooxidation under white light in $O_2$ atmosphere, similar to the technique recently used by Najafov *et al*. [9]. Exposing the crystal to doses of illumination in oxygen or air results in a gradual (and thus controlled) surface oxidation with increasing density of traps. This has been shown to systematically reduce the dark and photoconductivity of rubrene crystals [9]. In the experiment presented in Fig. S1 below, a pristine rubrene crystal was first measured and then exposed to a bright white light (100 mW·cm$^{-2}$) in air for a short period of time (~ 10 s), which has immediately led to a reduction of the dark and photoconductivity of the sample. After each additional dose of photooxidation, dark and photoconductivity were measured in high vacuum (with the gauge off). The resultant set of $\sigma_{PC}(G)$ curves reveals a remarkable result (Fig. S1). Each dose of photooxidation reduces the absolute value of photoconductivity and leads to a gradual evolution from the power exponent $\alpha = 1/3$ (corresponding to the surface-dominated photoconductivity in pristine crystals) to $\alpha = ½$ that corresponds to pure bulk photoconductivity. Interestingly, the drastic reduction of the absolute value of photoconductivity in this process (by almost two orders of magnitude) is irreversible, contrary to the case of the gauge-effect treated samples that can be recovered. However, the final state corresponding to fully oxidized surface (the lowest curve in Fig. S1) is identical in behavior to the gauge-effect treated samples that also show $\sigma_{PC} \propto G^{0.5}$ dependence with $\alpha = ½$ and a very small dark conductivity, $\sigma_0 \sim 50$ pS.

This additional experiment shows that the exact method of introducing surface defects is not important: we observe a similar transformation from α = 1/3 to α = ½ (accompanied by a drastic reduction of the absolute value of photoconductivity) after exposing the crystals to high-vacuum gauges, intentionally introducing microscopic mechanical scratches at the (*a,b*) facet between the contacts, as well as photooxidation.

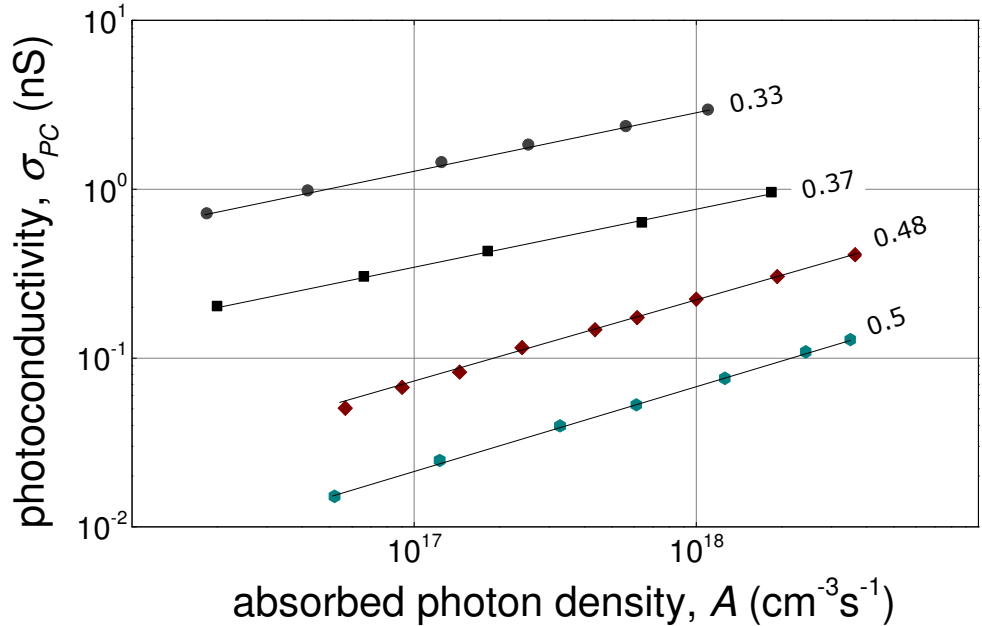

**Fig. S1. Excitation density dependence of photoconductivity in a gradually (photo)oxidized rubrene crystal.** For ease of comparison, only the intermediate range of excitation densities is shown. Photooxidation was performed by exposing the crystal to short doses of illumination in air, as described in sec. 3 of this Supplementary Information. Photoexcitation for $\sigma_{PC}$ measurements was achieved by illuminating the (*a,b*) facet of the crystal at a normal incidence with a monochromatic light (λ = 500 nm, penetration length $\alpha^{-1}$ = 1 μm) in high vacuum with the gauge off. As the density of oxygen-related traps near the surface of the crystal increases, $\sigma_{PC}(G) \propto G^\alpha$ dependence clearly shows a gradual transformation from α = 1/3 (surface photoconductivity) to α = ½ (bulk photoconductivity).

## 4. Comparison of the mechanisms of sub-linear photoconductivity in *rubrene* and *Si*.

It can be seen in Fig. S2a that the photocurrent response of rubrene crystals becomes sublinear already at very low excitation powers. Sublinearities are sometimes observed in organic or inorganic semiconductors and are often attributed to either an electron-hole ($np$) recombination (for $\alpha = 1/2$) or the cooperation of traps and recombination centers (for $\alpha \neq 1/2$). Sublinearity with $\alpha$ lower than ½ at elevated temperatures was modeled for an organic semiconductor in Ref. [10] assuming the diffusion dominant transport of photoexcited carriers and an exponential trap distribution. $\alpha = 1/3$ dependence at high illumination intensities observed in crystalline germanium can be expected when Auger recombination (that is, a non-radiative, three particle recombination, $nnp$ or $ppn$) is dominant, and carrier diffusion is negligible [11].

Given the recent progress in the growth of organic single crystals and the availability of samples with unprecedented chemical purity (such as rubrene), we do not expect that residual small concentration of traps in pristine crystals can be sufficient to result in the observed sublinearities in $\sigma_{PC}$.

We propose that when the equilibrium concentrations of excitons and charge carriers in our *cw* measurements become sufficiently high, an additional loss mechanism, given by the triplet-charge quenching term ($T \cdot n$) in the rate equations 1-3 formulated in the main text, becomes not only relevant, but dominant. As we show in the main text, this can ultimately lead to $\alpha = 1/3$ power dependence of photoconductivity. It must be noted that a regime with $\alpha = 1/3$ can, in principle, result from a third-order Auger carrier recombination. Such processes can occur, typically in indirect-gap semiconductors (for instance, Si or Ge), where radiative recombination of electrons and holes is forbidden, and a third carrier is necessary in order to carry away the excess energy and momentum [12,13,14]. This can typically be observed in Si or Ge at a very high concentration of photocarriers.

Indeed, as we show in Fig. S2b for crystalline Si(111), one can achieve a situation when a third-order (three-particle) mechanism of sublinearity in photoconductivity becomes apparent. This can only be observed when using a very high photoexcitation intensity (when the photocarrier concentration is high), or when the material is of exceptionally high purity and has a high carrier mobility, that is, when other recombination channels (involving bang-gap states) are suppressed. This third-order recombination mechanism is related to the Auger effect, in which the excess energy and momentum are transferred to a third carrier that gets excited to a higher energy level within the band. After such a three-particle interaction, the third carrier eventually losses its excess energy to thermal lattice vibrations.

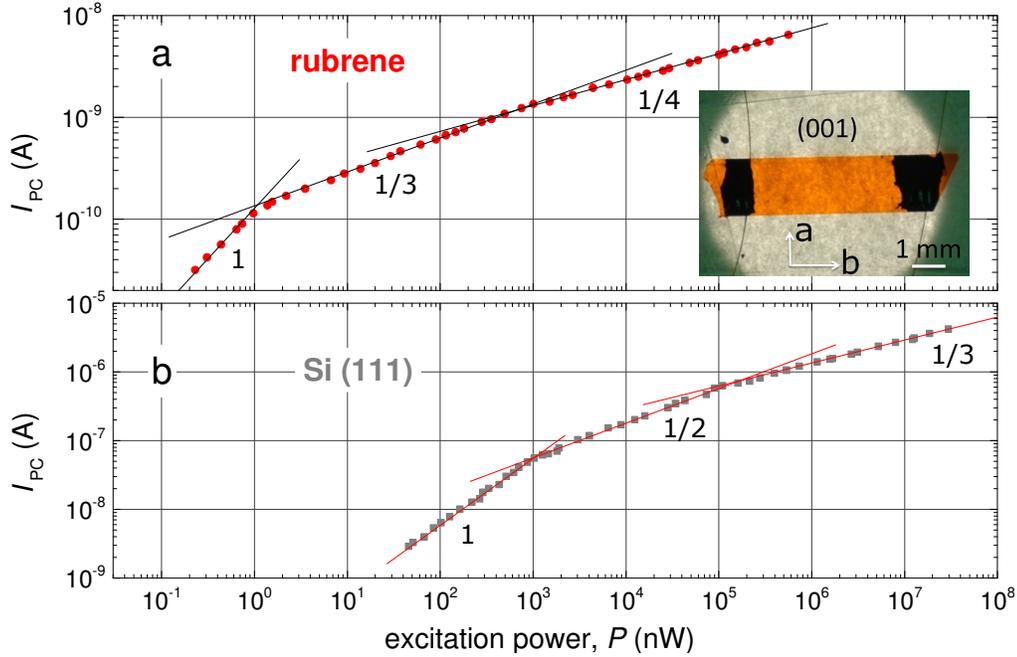

**Fig. S2. Excitation density dependence of photoconductivity in (a) pristine rubrene crystals, compared to similar dependence in (b) single-crystal Si(111).** **(a):** Photoexcitation achieved by illuminating the ($a,b$) facet of the rubrene crystal at a normal incidence with a monochromatic light ($\lambda =$ 500 nm, $\alpha^{-1} = 1$ μm) in high vacuum with the gauge off and $V = 1$ V applied between the contacts separated by 4 mm. **(b):** Photoexcitation achieved by illuminating the (111) facet of the high-resistivity Si sample ($\rho = 30$ kΩ cm) at a normal incidence with a monochromatic $cw$ laser light ($\lambda = 532$ nm, $\alpha^{-1} = 1.36$ μm) in high vacuum with the gauge off and $V = 1$ V applied between contacts separated by 5 mm. All experimental conditions, including the sample dimensions and excitation light penetration lengths, were intentionally chosen to be very similar for convenient direct comparison between the two materials. Note that the deviation from the linear growth in the case of pristine rubrene occurs at many orders of magnitude lower excitation power compared to Si, which suggests that third-order Auger recombination mechanism is an unlikely mechanism for rubrene, and a model based on a second-order $T \cdot n$ annihilation should be favored (see main text).

Our experiments show that in rubrene single crystals, the onset of power-law dependence with α = 1/3 occurs at a very low excitation intensity (6 orders of magnitude lower than that in Si), and this regime persists over a wide range of excitation powers (Fig. S2). One must consider the difference between these two materials with respect to the particle interactions and dynamics. While in Si (at room temperature) photon absorption leads to a direct creation of mobile electrons and holes, photon absorption in rubrene leads to a generation of excitonic states with relatively long lifetimes of a few nanoseconds for singlets and ~ 100 microseconds for triplets, and charge carrier generation in rubrene is a secondary process that occurs due to assisted dissociation of such long-lived excitons. This results in a situation when steady-state photoexcitation can produce relatively large populations of triplet excitons and charge carriers, and new decay channels, such as triplet-charge quenching, become dominant. This, however, is not expected in Si because of the absence of stable excitons at room temperature.

The fact that sublinearities in the power dependence of photoconductivity in Si occur at much greater excitation densities than in rubrene (Fig. S2), when the two systems are measured in similar experimental conditions, is consistent with the bulk vs. surface nature of photoconductivity in Si and pristine rubrene, respectively. In addition, the sequence of exponents in pristine rubrene is different from that in Si: in rubrene we have $\alpha = 1$, then $1/3$, and then $¼$, with the electron-hole recombination regime ($\alpha = 1/2$) missing. This indicates that $\alpha = 1/3$ regime in rubrene cannot be assigned to a three-particle Auger recombination, as in such a case it would have been preceded by the $\alpha = ½$ regime due to a more probable *np* recombination. This clearly shows that the mechanisms of sub-linearities in photoconductivity in molecular crystals are *fundamentally different* from those in inorganic semiconductors.

## 5. Refinement of the photoconductivity "oscillations" model (Ref. [7]).

The sub-linear excitation intensity dependence of photoconductivity observed in this work spans many orders of magnitude in excitation powers and covers the range typically used in laboratory studies of photophysical properties of organic materials. For instance, the regime described by the function $\sigma \propto G^{1/3}$ (the so-called $\alpha = 1/3$ regime) extends over three-four orders of magnitude in illumination intensity (Figs. 1 and 3b of the main text and Fig. S2a above). Therefore, it is very likely that most (if not any) existing measurements, in which the excitation density is varied, are affected by this dependence.

An important measurement that has been recently performed in rubrene concerns the modulations of photoconductivity with the polarization angle of light, $\sigma_{PC}(\theta)$ (the so-called "oscillations" in photoconductivity) [7]. Here, $\theta$ refers to an angle between the polarization of a monochromatic linearly polarized light normally incident on the (*a*,*b*) facet of rubrene crystal and the crystallographic *b*-axis of the crystal (the high-mobility axis). As usual, the photoconductivity in this experiment has been measured in a co-planar contact geometry at the (*a*,*b*) facet of the crystal. It has been found that in high-purity pristine rubrene crystals, the contrast of photoconductivity modulations, $\eta_\sigma$, is much smaller than the corresponding contrast of the absorption coefficient, $\eta_\alpha$ (Fig. S3 below, reproduced from Ref. [7]). Here, again, the absorption coefficient of rubrene, **α**, should not be confused with the power exponent α in σ(*G*) dependence. The contrasts in $\sigma_{PC}$ and **α** are defined as:

$$\eta_\sigma \equiv (\sigma_b - \sigma_a)/\sigma_b \quad \text{and} \quad \eta_\alpha \equiv (\alpha_b - \alpha_a)/\alpha_b, \quad \text{respectively.}$$

For example, in pristine rubrene crystals, the range of σ-contrast values obtained in a large number of samples at 495-nm excitation is $\eta_\sigma$ = 3 - 12 %, with more typical values of 5 - 9 % observed in high-purity crystals, while the corresponding modulations of the absorption coefficient **α**(θ) are much greater, $\eta_\alpha$ = 42 % (Fig. S3, see also Ref. [7]).

This effect can only be rationalized by assuming a long-range diffusion of photogenerated excitons to the surface of the crystal, where they dissociate and generate surface photocurrent. In this model, photon absorption initially results in a *Beer-Lambert* distribution of singlet excitons below the (*a*,*b*) surface of the crystal, shown as a hatched area in Fig. S3b. Singlet excitons in rubrene rapidly undergo fission into triplets that are forbidden from a radiative recombination and therefore have a long lifetime ($\tau_T$ ~ 100 µs) [2]. Singlet fission in rubrene crystals has been confirmed in several recent experiments (refs. [9], [11], [13], [16], [23] and [24] of the main text). The rather low integral *PL* quantum yield in rubrene crystals (only about 1-2 % at low to moderate photoexcitation intensities) is consistent with singlet fission occurring in this material. Such *PL* quantum efficiency is still greater than that in tetracene or pentacene, which are both known to be singlet fission materials. This however can be understood based on the fact that in tetracene and pentacene, the first excited triplet energy $T_1$ is smaller than one half of the singlet energy $S_1$, and thus triplet fusion requires a thermal activation. On the contrary, in rubrene, triplet energy is believed to be very close to one half of the singlet energy, and thus fusion process is rather efficient at room temperature, leading to a recovery of some singlets. This makes *PL* quantum yield in rubrene higher in *cw* measurements. The triplet excitons in rubrene can migrate throughout the crystal over the distances much greater than the typical diffusion range for singlet excitons and dissociate on defects (including the natural boundaries of crystals – their surfaces), thus generating photoconductivity.

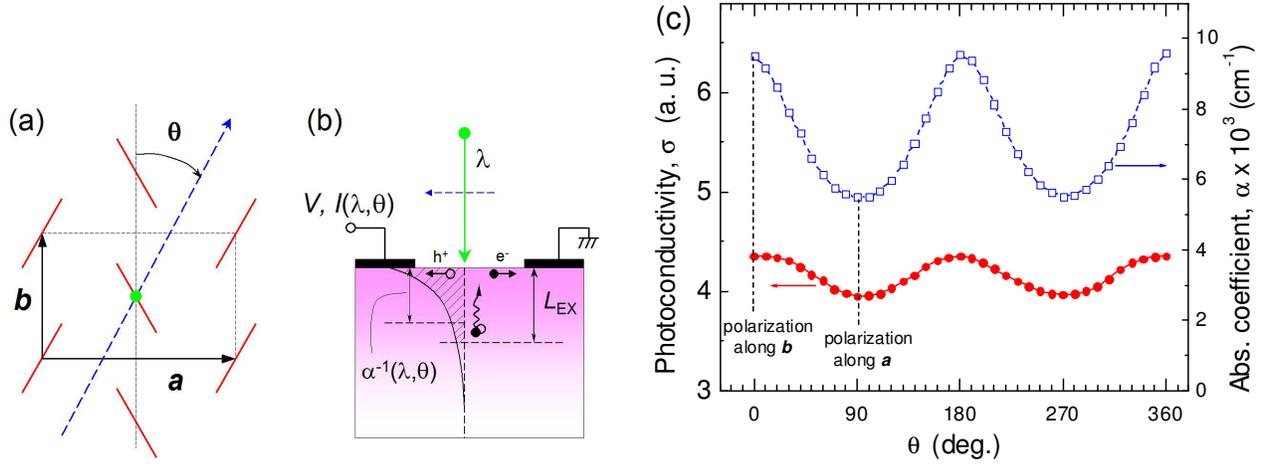

**Fig. S3 (reproduced from Ref. [7]).** (a) Molecular packing in the (*a*,*b*) plane of rubrene: long molecular axes are shown by the red segments, blue dashed arrow defines the polarization of a normally incident light with an angle $\theta$ with respect to the *b*-axis. (b) A schematic "side view" of the device showing electrical contacts at the (*a*,*b*)-facet of the crystal probing the photocurrent response of the sample, $I(\lambda,\theta)$, to incident light (green arrow). An exponential profile of the density of photogenerated excitons in the crystal defined by the light penetration length, $\alpha^{-1}(\lambda,\theta)$, and an exciton diffusion range, $L_{EX}$, are shown. (c) Experimental dependence of the photocurrent on polarization angle for a 495-nm excitation (solid red circles), measured in a pristine rubrene crystal with thickness $d = 1$ mm $\gg \alpha^{-1}(495 \text{ nm}) \approx 1$ µm (that is, absorbing all the incident photons). Angular dependence of the absorption coefficient, $\alpha(\theta)$, for 495-nm excitation in rubrene (open blue squares).

This effect has been modeled by Najafov *et al.* by taking the integral that counts the total number of triplet excitons reaching the top surface of the crystal (for details, see Ref. [7]). This integral is reproduced here:

$$\sigma_{PC}(\theta, \lambda) = \sigma_{surf} = \gamma_0 \chi_0 \Phi_0 \alpha \int_0^\infty exp(-\alpha \cdot x) \cdot exp(-x/L_{EX}) \cdot dx = \gamma_0 \chi_0 \Phi_0 \cdot \frac{\alpha L_{EX}}{\alpha L_{EX} + 1} \quad (S1)$$

In this integral, $\alpha \cdot \Phi_0 \cdot \exp(-\alpha x) \cdot dx$ is the number of photons absorbed in a layer of the crystal below the (*a*,*b*) surface from $x$ to $x + dx$ ($x$ represents the depth into the crystal with $x = 0$ corresponding to the top (*a*,*b*) surface of the crystal). The second exponent, $\exp(-x/L_{EX})$, is related to the probability for a triplet exciton generated at the depth $x$ to reach the top surface of the crystal. According to this model, the photoconductivity oscillations would then be described by the contrast:

$$\eta_\sigma \equiv \frac{\sigma_b - \sigma_a}{\sigma_b} = 1 - \frac{\alpha_a}{\alpha_b} \cdot \frac{\alpha_b L_{EX} + 1}{\alpha_a L_{EX} + 1} \tag{S2}$$

In this equation, the triplet exciton diffusion length, $L_{EX}$, is the only fitting parameter, with $\alpha_{a,b}(\lambda)$ being the experimental absorption coefficients measured for polarizations along *a* and *b* axes of the crystal.

In light of the new knowledge of nonlinear surface photoconductivity described in the current work (see the main text) the above model of Najafov *et al.* must be adjusted. According to the model of the regime with power exponent α = 1/3 (this regime occurs at most commonly used photoexcitation intensities), photoconductivity in this regime can be derived from the set of rate equations given in the main text (eqs. 1-3) under the assumption that the triplet fusion term in eqs. 1 and 2 (proportional to $T^2$) is not yet dominant, the two other terms ($T/\tau_T$ and $\beta T$) are also small and can be neglected, because the triplet lifetime $\tau_T$ is long, and the quantum efficiency of triplet dissociation at the surface, $\beta$, is relatively small. This leaves only two dominant terms in eq. 2: the triplet generation via singlet fission ($2f_S S/\tau_S$) and the triplet-charge quenching at the surface ($\gamma_a T \cdot n$). Thus, in this regime, the set of rate equations 1-3 reduces to the following:

$$\frac{dS}{dt} = G - \frac{S}{\tau_S} \tag{S3}$$

$$\frac{dT}{dt} = 2f_S \frac{S}{\tau_S} - \gamma_a T n \tag{S4}$$

$$\frac{dn}{dt} = \beta T - \gamma_n n^2 \tag{S5}$$

These equations lead to the following steady-state ($dS/dt = dT/dt = dn/dt = 0$) solution for the densities of singlets, triplets and photocarriers: $S \propto G$, $T \cdot n \propto G$, and $n \propto G^{1/3}$.

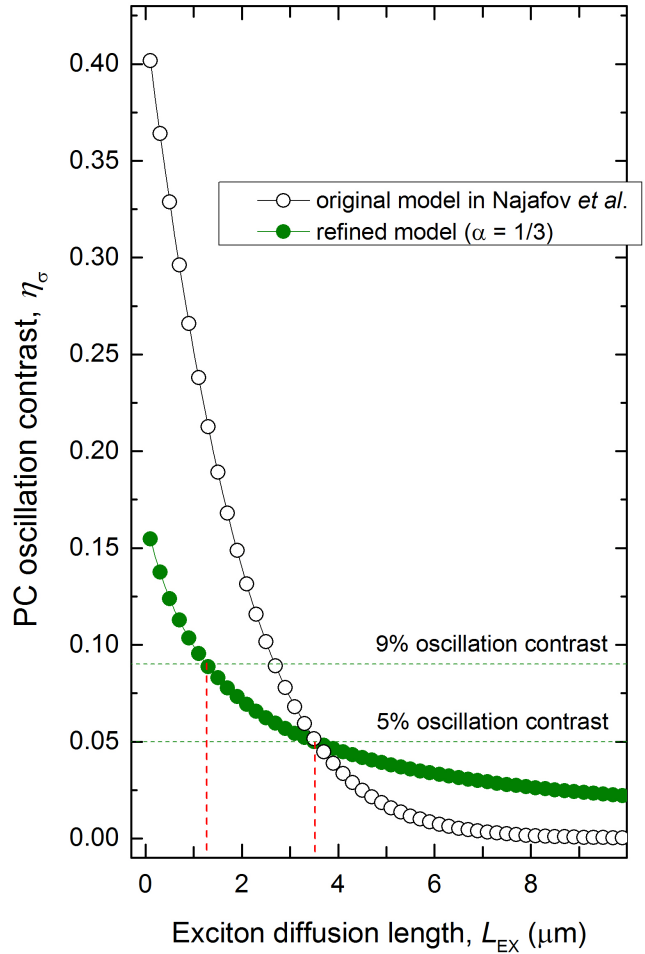

**Fig. S4.** The contrast of photocurrent modulations, $\eta_\sigma$, plotted as a function of exciton diffusion length, $L_{EX}$, calculated by Najafov et al. [7] (black open symbols), and that calculated by using the refined Eq. S8 that accounts for the sublinear photoconductivity in the regime $\sigma_{PC} \propto G^{1/3}$ (green solid symbols). Both curves were obtained using the experimental values of the absorption coefficient $\alpha$ for 495-nm excitation polarized along $b$ and $a$ axes of rubrene ($\alpha_a = 5.5 \times 10^3$ cm$^{-1}$, and $\alpha_b = 9.5 \times 10^3$ cm$^{-1}$). The horizontal dashed lines mark the range of typical experimental values for the oscillation contrast in $\sigma$, $\eta_\sigma = 5$ -9 %. It can be seen that in the refined model, the exciton diffusion length is quite substantial, $L_{EX} > 1$ μm.

Note that eq. S4 for the steady-state density of triplets $T$ has two terms: 1$^{st}$ is the source of triplets via singlet fission, and 2$^{nd}$ is the most relevant (in this regime) channel of triplet decay via triplet-charge quenching. The combination of these competing processes is what determines the steady-state triplet population, $T$, which in turn governs the photocarrier production described by eq. S5. The source of triplets via singlet fission plays an equivalent role as the integral used by Najafov *et al.* in their original modeling of photocurrent oscillations (the integral in eq. S1 above) [7]. The only difference between that integral and the term $2f_S S/\tau_S$ in eq. S4 is that the integral calculates the portion of triplets that reach the top surface of the crystal via diffusion characterized by exciton diffusion length, $L_{EX}$, while the 1$^{st}$ term in eq. S4 refers to the overall triplet population.

Thus, in order to make the necessary refinement of Najafov's model, one has to take into account both the triplet source and the triplet decay channels, so that a steady-state population of triplets, $T$, is used, instead of just the source of triplets. Such an adjustment of the model can be easily done by taking the expression on the right-hand side of eq. S1 (the result of the integration) and using it as the source of triplets in eq. S4:

$$2f_S S/\tau_S \quad \Rightarrow \quad \gamma_0 \chi_0 \Phi_0 \cdot \frac{\alpha L_{EX}}{\alpha L_{EX}+1} \tag{S6}$$

After this, instead of $T \cdot n \propto G$ and $n \propto G^{1/3}$, we get:

$$T \cdot n \propto \frac{\alpha L_{EX}}{\alpha L_{EX}+1} \text{ (from eq. S4), and } n \propto \left(\frac{\alpha L_{EX}}{\alpha L_{EX}+1}\right)^{1/3} \text{ (from eq. S5), respectively, leading to}$$

the photoconductivity,

$$\sigma_{PC} \equiv en\mu \propto \left(\frac{\alpha L_{EX}}{\alpha L_{EX}+1}\right)^{1/3} \tag{S7}$$

The refined expression for the contrast of photoconductivity oscillations will then take a form:

$$\eta_\sigma \equiv \frac{\sigma_b - \sigma_a}{\sigma_b} = 1 - \left(\frac{\alpha_a}{\alpha_b} \cdot \frac{\alpha_b L_{EX}+1}{\alpha_a L_{EX}+1}\right)^{1/3} \tag{S8}$$

This expression (instead of the original one in (S2)) should be used to model the photoconductivity oscillations. Figure S.4 performs such a modeling using both the old formula from Najafov *et al.* [7] (eq. S2) and the new formula (eq. S8). The refined model (green solid symbols) leads to somewhat smaller exciton diffusion lengths for the oscillation contrasts > 5%, while it gives longer $L_{EX}$ for smaller contrasts. Nevertheless, even at a very large oscillation contrast of 10 % (which is typically a sign of insufficiently purified crystals), the exciton diffusion length predicted by the refined model is ~ 1 µm.

Finally, we would like to emphasize that the co-planar device geometry (electrical contacts separated by a large distance of 1-3 mm, deposited at the top surface of the crystal), as well as macroscopically thick crystals used in Najafov *et al*. [7] and in this work by Irkhin *et al*. ensure that there are no experimental artifacts related to the unaccounted effects of interference and exciton quenching at metal/organic interfaces that are very common in thickness-dependent PL quenching and photo-current excitation spectroscopy measurements in sandwiched thin films. An excellent review of various artifacts that can lead to erroneous measurements of exciton diffusion length in thin-film devices can be found in the recent paper by R. R. Lunt *et al*. [15].

**References:**

[1] R. W. I. de Boer, M. E. Gershenson, A. F. Morpurgo, and V. Podzorov, Phys. Status Solidi A **201**, 1302 (2004).

[2] P. Irkhin, A. Ryasnyanskiy, M. Koehler, and I. Biaggio, Phys. Rev. B **86**, 085143 (2012).

[3] V. Podzorov, V. M. Pudalov, and M. E. Gershenson, Appl. Phys. Lett. **85**, 6039 (2004).

[4] V. Nardello, M.-J. Marti, C. Pierlot, and J.-M. Aubry, J. of Chem. Education **76**, 1285 (1999).

[5] D. D. T. Mastrogiovanni, J. Mayer, A. S. Wan, A. Vishnyakov, A. V. Neimark, V. Podzorov, L. C. Feldman, and E. Garfunkel, Sci. Reports **4,** 4753 (2014).

[6] Y. Wakabayashi, J. Takeya, and T. Kimura, Phys. Rev. Lett. **104**, 066103 (2010).

[7] H. Najafov, B. Lee, Q. Zhou, L. C. Feldman, and V. Podzorov, Nature Mater. **9**, 938 (2010).

[8] V. Podzorov, E. Menard, S. Pereversev, B. Yakshinsky, T. Madey, J. A. Rogers, and M. E. Gershenson, Appl. Phys. Lett. **87**, 093505 (2005).

[9] H. Najafov, D. Mastrogiovanni, E. Garfunkel, L. C. Feldman, and V. Podzorov, Adv. Mater. **23**, 981 (2011).

[10] Y. Aoyagi, K. Masuda, and S. Namba, J. Appl. Phys., **43**, 249 (1972).

[11] R. E. Wagner and A. Mandelis, Semicond. Sci. Technol. **11**, 300 (1996).

[12] M. J. Kerr and A. Cuevas, J. Appl. Phys. **91**, 2473 (2002).

[13] A. Richter, S. W. Glunz, F. Werner, J. Schmidt, and A. Cuevas, Phys. Rev. B **86,** 165202 (2012).

[14] E. Yablonovitch and T. Gmitter, Appl. Phys. Lett. **49**, 587 (1986).

[15] R. R. Lunt, S. R. Forrest *et al*. *J. Appl. Phys.* **105**, 053711 (2009).